\documentclass{article}

\usepackage{microtype}
\usepackage{graphicx}
\usepackage{subfigure}
\usepackage{booktabs} 
\usepackage{multirow}

\usepackage{hyperref}

\usepackage[accepted]{icml2019}

\icmltitlerunning{Zero-shot Learning and Knowledge Transfer in Music Classification and Tagging}

\begin{document}

\twocolumn[
\icmltitle{Zero-shot Learning and Knowledge Transfer in Music Classification and Tagging}




\begin{icmlauthorlist}
\icmlauthor{Jeong Choi} {kaist}
\icmlauthor{Jongpil Lee} {kaist}
\icmlauthor{Jiyoung Park} {naver}
\icmlauthor{Juhan Nam} {kaist}
\end{icmlauthorlist}

\icmlaffiliation{kaist}{Graduate School of Culture Technology, Korea Advanced Institute of Science and Technology (KAIST), Daejeon, South Korea}
\icmlaffiliation{naver}{NAVER Corp., South Korea}
\icmlcorrespondingauthor{Juhan Nam}{juhannam@kaist.ac.kr}
\icmlkeywords{Machine Learning, ICML}

\vskip 0.3in
]



\printAffiliationsAndNotice{}  

\begin{abstract}
Music classification and tagging is conducted through categorical supervised learning with a fixed set of labels. In principle, this cannot make predictions on unseen labels. Zero-shot learning is an approach to solve the problem by using side information about the semantic labels. We recently investigated this concept of zero-shot learning in music classification and tagging task by projecting both audio and label space on a single semantic space. In this work, we extend the work to verify the generalization ability of zero-shot learning model by conducting knowledge transfer to different music corpora. 

\end{abstract}


\section{Introduction}
\label{introduction}
Zero-shot learning is a paradigm of machine learning to overcome the limitations of categorical supervised learning that can predict only a fixed number of classes. By leveraging additional side information regarding classes, zero-shot learning models can discover relationship among any arbitrary classes with regard to the given task, enabling inference towards classes that are unseen during training.
We recently applied this approach to the music domain in attempts to allow prediction of novel music genres or retrieval of songs by a query word that users arbitrarily choose \cite{zsl_music}. 
By splitting tag labels into seen/unseen groups, we verified that the concept of knowledge transfer is possible within a dataset. 
In this study, we extend the evaluation of zero-shot embedding model to investigate how learned semantic relationships are transferred not only within a single dataset but also across different music corpora.


\begin{figure}[ht]
\begin{center}
\centerline{\includegraphics[width=\columnwidth]{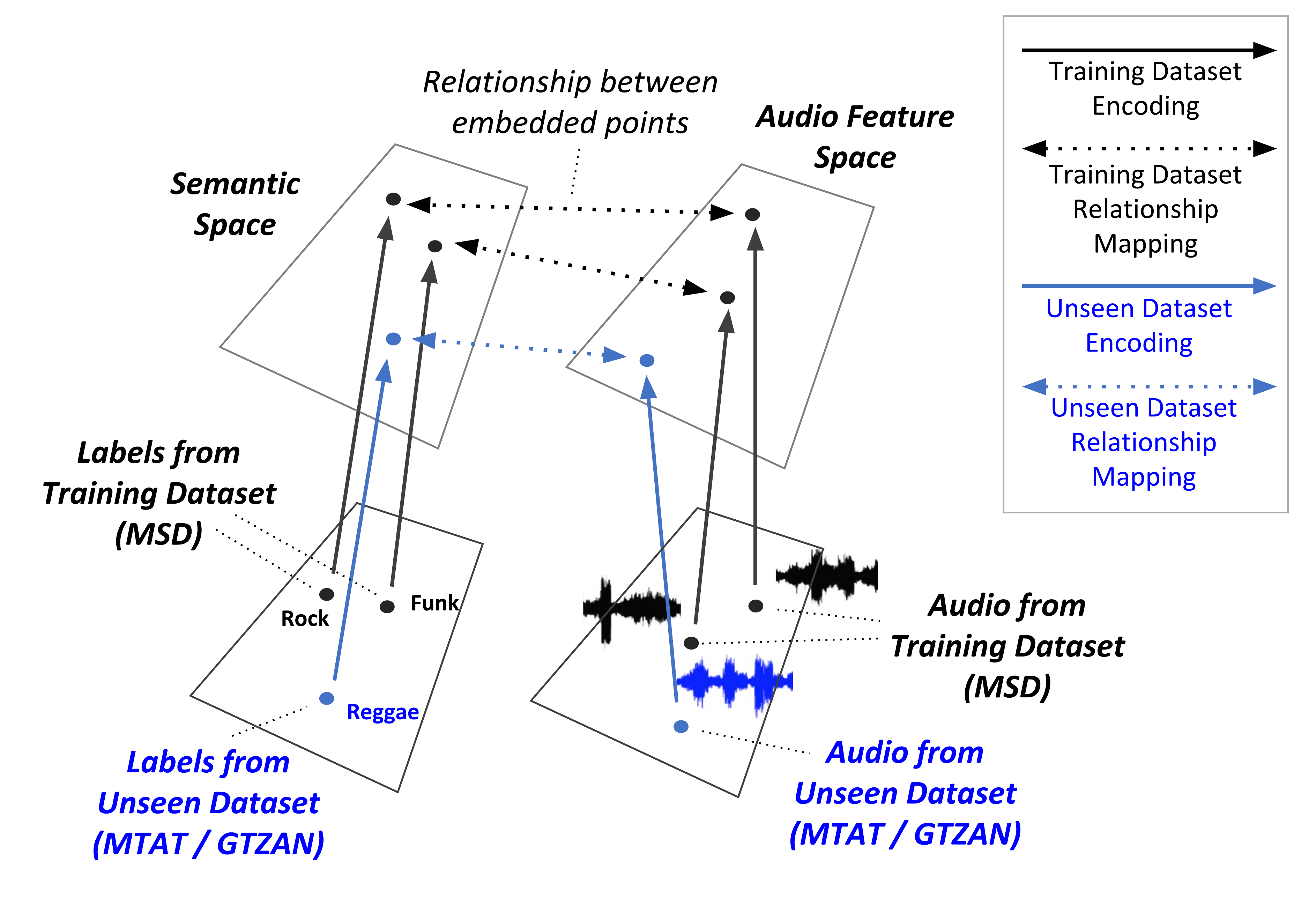}}
\caption{Overview of zero-shot learning and knowledge transfer applied to music domain.}
\label{zsl_model}
\end{center}
\vskip -0.3in
\end{figure}

\section{Zero-Shot Learning for Music}
Figure \ref{zsl_model} illustrates the overview of zero-shot learning in the music domain. The semantic space is constructed using side information about the labels such as instrument annotations that describe music genres or a pre-trained word vector space such as GloVe \cite{glove}. The audio feature space is extracted via a convolutional neural network (CNN) that takes mel-spectrogram as input. The audio feature extraction and the mapping between the semantic space and audio feature space are learned using a Siamese-style network with the triplet loss, following the previous works \cite{devise, park2017representation}. Once the networks are trained, audio examples from the test set can be mapped to a point in the semantic space whose nearest label can be unseen ones during the training phase.

\section{Knowledge Transfer}
We conducted the evaluation of the zero-shot learning model within the same music corpus in our previous work \cite{zsl_music}. Since each of datasets has different audio characteristics and annotation criteria, evaluating the trained zero-shot model on different datasets can provide an insight into its generalization ability. 

To this end, we train the zero-shot model on the Million Song Dataset (MSD) \cite{msd} with the Last.fm tag annotations and then evaluate the trained model on MagnaTagATune (MTAT) \cite{mtat} and GTZAN \cite{gtzan} by directly feeding the test set of tracks to the network. We measure the classification performance and compare it with a referenced model.

\begin{table}[t]
\caption{Classification results. Tag retrieval AUC is reported for the MTAT dataset and genre annotation accuracy is reported for the GTZAN dataset.} 

\label{tab:table1}
\vskip 0.15in
\begin{center}
\begin{small}
\begin{sc}

\resizebox{\columnwidth}{!}{\begin{tabular}{@{}ccc@{}}
\toprule
Model
& \begin{tabular}[c]{@{}c@{}}
MTAT\\ 
(Ret. AUC)
\end{tabular} 
& \begin{tabular}[c]{@{}c@{}}
GTZAN\\ 
(Anno. Acc.)
\end{tabular} \\ 
\midrule
\begin{tabular}[c]{@{}c@{}} Zero-Shot learning model \\ (MSD-GloVe) \end{tabular} 
&  0.7390 & 0.7310 \\
\midrule
\begin{tabular}[c]{@{}c@{}} Classification model \\ \cite{lee2017multi} \end{tabular} 
 & 0.9021  &  0.7203 \\ 
\bottomrule
\end{tabular}}

\end{sc}
\end{small}
\end{center}
\vskip -0.1in
\end{table}

\section{Experiments}

The zero-shot learning model is composed of audio and word branches. The audio branch consists of CNN layers and the word branch contains a linear projection layer on top of the side information lookup table built from a pre-trained general word semantic space. The two branches are then jointly trained with a max-margin hinge loss. We train the zero-shot model on MSD using 1,126 tags that are present in the side information lookup table. 

We chose the GloVe embedding as our baseline side information for its large vocabulary that facilitates adaptation to unseen datasets \cite{glove}. We utilized a pre-trained GloVe model available online. It contains 19 million vocabularies with 300 dimensional embedding trained from documents in Common Crawl data. We then evaluated the model on MTAT and GTZAN. For the MTAT dataset, we followed the evaluation setup of using 50 most frequent tags \cite{dieleman2014end}. 7 tags that are not present in GloVe vocabulary were omitted, resulting in total of 43 tags. For the GTZAN dataset, we used a fault-filtered version split \cite{kereliuk2015deep} and all 10 genre tags are used as all of them are presented in the GloVe dictionary. The evaluation is conducted on the test tracks of the referenced split for each dataset.

\section{Results}

The results of knowledge transfer are presented in Table \ref{tab:table1}. On GTZAN, the zero-shot learning model shows a significant improvement, even surpassing the performance of referenced supervised model which was supervised with the training set of GTZAN. On MTAT, on the contrary, the performance score is relatively low. We speculate that this is because the property of each dataset is different. GTZAN is comparatively more similar to MSD in terms of genre distribution of tracks and their general audio characteristics than MTAT. 

We also present case studies of knowledge transfer evaluation. We investigated the predicted tags both from the label candidates of the test dataset and from the tag pool of the training dataset. 
The predictions on MTAT tracks and GTZAN tracks are shown in Table \ref{tab:ret_word_mtat} and Table \ref{tab:ret_word_gtzan}, respectively. From the results, we can see that that the model is able to predict labels that are semantically close to the ground truth labels even though no association between tracks and labels was informed to the model.

\section{Conclusion and Future Work}

We examined how well the knowledge is transferred across different music classification and tagging datasets via the zero-shot embedding space. For future work, we will explore the side information more thoroughly. 
For example, a word embedding reflecting more music-specific semantic relationship (e.g., trained with song review articles) would improve the performance of the model.

\begin{table}[t]
\caption{Comparison of label predictions for MTAT tracks on the original MTAT tags and 1,126 MSD Last.fm tags.}
\label{tab:ret_word_mtat}
\vskip 0.15in
\begin{center}
\begin{small}
\begin{sc}

\resizebox{\columnwidth}{!}{%
\begin{tabular}{@{}cccc@{}}
\toprule
Track ID
& \begin{tabular}[c]{@{}c@{}} 
 Ground Truth \\ 
 (MTAT)
 \end{tabular}  
& \begin{tabular}[c]{@{}c@{}} Predictions out of \\  MTAT tags \\ (threshold = 0.6) \\ \end{tabular}          & \begin{tabular}[c]{@{}c@{}} Predictions out of \\ 1,126 MSD tags \\ (threshold = 0.7) \end{tabular} \\ \midrule
 8900  
 & \begin{tabular}[c]{@{}c@{}} 
 slow \\ 
 ambient \\ 
 synth 
 \end{tabular}
 & \begin{tabular}[c]{@{}c@{}}
 ambient \\ 
 piano \\ 
 electronic 
 \end{tabular}
 & \begin{tabular}[c]{@{}c@{}} 
 instrumental \\  
 chillout / ambient \\ 
 soundtrack / jazz / piano \\ 
 relaxing\\
 \end{tabular}
 \\ 
 \midrule
11308
& \begin{tabular}[c]{@{}c@{}} 
 slow / techno\\ 
 electronic / beat \\
 synth / weird \\
\end{tabular}
& \begin{tabular}[c]{@{}c@{}} 
 electronic \\ 
 techno \\ 
 dance 
\end{tabular}
& \begin{tabular}[c]{@{}c@{}} 
electronic / techno \\ 
electro / electronica \\ 
downtempo / dub \\ 
bassline / chillout \\
experimental \\

\end{tabular}
\\ 
 
\bottomrule
\end{tabular}%
}

\end{sc}
\end{small}
\end{center}
\vskip -0.1in
\end{table}

\begin{table}
\caption{Comparison of label predictions for
GTZAN tracks on the original GTZAN genres and 1,126 MSD Last.fm tags.}
\label{tab:ret_word_gtzan}
\begin{center}
\begin{small}
\begin{sc}

\resizebox{\columnwidth}{!}{%
\begin{tabular}{@{}cccc@{}}
\toprule
Track ID
& 
\begin{tabular}[c]{@{}c@{}} 
Ground Truth \\ 
(GTZAN)
\end{tabular}          
& \begin{tabular}[c]{@{}c@{}} Top 1 Prediction out of \\  GTZAN genres \end{tabular}          & \begin{tabular}[c]{@{}c@{}} Predictions out of \\ 1,126 MSD tags \\ (threshold = 0.6) \end{tabular} \\ \midrule
 560            & jazz        & jazz 
                                & \begin{tabular}[c]{@{}c@{}} jazz / blues \\ instrumental  \\  
                                swing / smoothjazz \\ piano  \end{tabular}
                                  \\ 
                                  \midrule
    600            & reggae        & reggae 
                                & \begin{tabular}[c]{@{}c@{}} reggae / dub / ska  \\  
                                roots / dancehall \\
                                rootsreggae \end{tabular}
                                  \\ 
 
\bottomrule
\end{tabular}%
}

\end{sc}
\end{small}
\end{center}
\vskip -0.1in
\end{table}

\bibliography{reference}
\bibliographystyle{icml2019}

\end{document}